\documentclass[acmsmall]{acmart}

\AtBeginDocument{%
  \providecommand\BibTeX{{%
    \normalfont B\kern-0.5em{\scshape i\kern-0.25em b}\kern-0.8em\TeX}}}

\setcopyright{acmcopyright}
\copyrightyear{2018}
\acmYear{2018}
\acmDOI{XXXXXXX.XXXXXXX}

\acmJournal{TOSEM}
\acmVolume{37}
\acmNumber{4}
\acmArticle{111}
\acmMonth{8}

\usepackage{multirow}
\usepackage{rotating}

\usepackage[many]{tcolorbox}    
\newtcolorbox{mybox}[1]{%
    tikznode boxed title,
    enhanced,
    arc=0mm,
    interior style={white},
    attach boxed title to top center= {yshift=-\tcboxedtitleheight/2},
    fonttitle=\bfseries,
    colbacktitle=white,coltitle=black,
    boxed title style={size=normal,colframe=white,boxrule=0pt},
    title={#1}}

\begin{document}

\title{Bug Analysis in Jupyter Notebook Projects: An Empirical Study}

\author{Taijara Loiola de Santana}
\email{emailTAIJARA}
\orcid{1234-5678-9012}
\affiliation{%
  \institution{Federal University of Bahia, Institute of Computing (IC-UFBA)}
  \streetaddress{rua}
  \city{Salvador}
  \state{Bahia}
  \country{Brazil}
  \postcode{CEP}
}
\author{Paulo Anselmo da Mota Silveira Neto}
\email{paulo.motant@ufrpe.br}
\affiliation{%
  \institution{Federal University Rural of Pernambuco (UFRPE)}
  \streetaddress{Rua Dom Manuel de Medeiros, s/n}
  \city{Recife}
  \state{Pernambuco}
  \country{Brazil}
  \postcode{52171-900}
}

\author{Eduardo Santana de Almeida}
\email{email}
\affiliation{%
  \institution{Federal University of Bahia, Institute of Computing (IC-UFBA)}
  \streetaddress{Rua}
  \city{Salvador}
  \state{Bahia}
  \country{Brazil}
  \postcode{cep}
}

\author{Iftekhar Ahmed}
\email{email}
\affiliation{%
  \institution{University of California, Irvine}
  \streetaddress{Rua}
  \city{Salvador}
  \state{Bahia}
  \country{USA}
  \postcode{cep}
}

\renewcommand{\shortauthors}{Trovato and Tobin, et al.}

\begin{abstract}
Computational notebooks, such as Jupyter, have been widely adopted by data scientists to write code for analyzing and visualizing data. Despite their growing adoption and popularity, there has been no thorough study to understand Jupyter development challenges from the practitioners’ point of view. This paper presents a systematic study of bugs and challenges that Jupyter practitioners face through a large-scale empirical investigation. We mined 14,740 commits from 105 GitHub open-source projects with Jupyter notebook code. Next, we analyzed 30,416 Stack Overflow posts which gave us insights into bugs that practitioners face when developing Jupyter notebook projects. Finally, we conducted nineteen interviews with data scientists to uncover more details about Jupyter bugs and to gain insight into Jupyter developers' challenges. We propose a bug taxonomy for Jupyter projects based on our results. We also highlight bug categories, their root causes, and the challenges that Jupyter practitioners face. 
\end{abstract}



\maketitle

\section{Introduction}

Due to the increased availability of data and computing resources over the past few years, data science and analytics have become important areas of investigation \cite{DBLP:conf/icse/BegelZ14}. Data science is an emerging field that combines mathematics, statistics, computer science, and domain knowledge to derive insights from data \cite{10.1145/2500499, 10.1145/3324884.3416543}. Data analytics is the multidisciplinary science of quantitatively and qualitatively examining data to draw new conclusions or insights (exploratory or predictive) or for extracting and proving (confirmatory or fact-based) hypotheses about that information for decision-making and action \cite{10.1145/3076253}.

Jupyter, a free, open-source web application that allows users to write documents composed of text, equations, visualizations, and code snippets and their execution results \cite{DBLP:conf/icse/WangLZ20}, has become the most widely-used system for exploring and analyzing data \cite{DBLP:conf/msr/PimentelMBF19, DBLP:conf/vl/KoenzenES20}. Data analysts use computational notebooks to write and refine code to understand unfamiliar data, test hypotheses, and build models \cite{DBLP:conf/chi/HeadHBDD19}.

Even with the benefits and growing popularity of Jupyter Notebooks, it has presented some problems. Wang et al. \cite{DBLP:conf/icse/WangLZ20} analyzed a sample of 1982 Jupyter Notebooks and found that they contain code with poorly respect to the Python style conventions, with unused variables which are defined but never referenced, and accessing deprecated functions. Pimentel et al. \cite{DBLP:conf/msr/PimentelMBF19} investigated the reproducibility aspects of real notebooks using a corpus consisting of 1.159.166 unique notebooks collected from 264.023 GitHub repositories. Out of 863.878 attempted executions of valid notebooks (i.e, notebooks with defined Python version and execution order), only 24.11\% executed without errors. 

Other recent studies \cite{DBLP:conf/chi/HeadHBDD19, DBLP:conf/chi/ChattopadhyayPH20} have identified additional problems related to \textit{name-value inconsistency} where the name and the value of a variable do not match \cite{DBLP:journals/corr/abs-2112-06186} and \textit{dependencies}, which around 94\% of notebooks do not formally state or document dependencies \cite{DBLP:conf/icse/WangLZ21}. Data analysts also have called their code \textit{ad hoc, experimental, and throw-away code} \cite{DBLP:journals/tvcg/KandelPHH12}, besides describing notebooks as \textit{messy} \cite{DBLP:conf/chi/KeryRAJM18, DBLP:conf/chi/RuleTH18}, containing \textit{ugly code} and \textit{dirty tricks} in need of \textit{cleaning} and \textit{polishing} \cite{DBLP:conf/chi/RuleTH18}. 

In addition, with the popularity of the area (Glassdoor ranks data science as the \#3 job in America for 2022\footnote{50 Best Jobs in America for 2022 - https://www.glassdoor.com/List/Best-Jobs-in-America-LST\_KQ0,20.htm}), and the serious consequences that a bug can bring in a data science project (UK lost nearly 16,000 COVID-19 cases by exceeding spreadsheet data limit\footnote{Thousands of coronavirus cases were not reported for days in the UK because officials exceeded the data limit on their Excel spreadsheet - https://www.businessinsider.com/uk-missed-16000-coronavirus-cases-due-to-spreadsheet-failure-2020-10}), analyzing and improving Jupyter notebook projects have potentially relevant impact.  

This paper presents the first comprehensive study of bugs in Jupyter notebook projects and the challenges that data scientists face in practice. Analyzing historical bugs that occurred in a system is an important step to reduce bugs \cite{DBLP:conf/issre/ThungWLJ12}. It can provide relevant knowledge to develop new tools for bug detection, triage bug reports, locate likely bug spots, suggest possible fixes, and help to monitor and improve quality along the development process. 

The software engineering community has conducted a number of studies that investigate bugs in different domains \cite{DBLP:conf/issre/ThungWLJ12, 10.1145/3213846.3213866, 10.1145/3338906.3338955, DBLP:conf/icse/GarciaF0AXC20, 10.1145/3377811.3380409, DBLP:conf/icse/Makhshari021, 10.1145/3468264.3468559}. However, despite these efforts, the characteristics of bugs in Jupyter notebook projects have never been systematically studied. The community has also stressed "the strong need to analyze the quality of the notebooks" \cite{DBLP:conf/icse/WangLZ20, DBLP:conf/chi/ChattopadhyayPH20, DBLP:conf/icse/WangLZ21} to improve the quality and reliability of the code. 

To better understand bugs that appear in Jupyter notebook projects, we followed three steps. First, we initially mined 14,740 commits from 105 GitHub open source repositories with Jupyter notebook code. Next, we analyzed 30,416 Stack Overflow posts which gave us insights into bugs that software developers face when developing Jupyter notebook projects. Finally, we conducted semi-structured interviews with 19 data scientists to validate the findings identified in the previous steps and understand how these bugs impact the daily life of data scientists working with Jupyter notebook projects.

Our study has led to multiple findings. In particular, we identify eight classes of bugs, ten types of root causes, the frequent impact of bugs, and a taxonomy which can help practitioners understand the nature of bugs, and define possible strategies to mitigate them. These findings can be used by researchers and practitioners to gain a better understanding of bugs in Jupyter notebook projects and point out new direction for future research.

Overall, the paper makes the following contributions: 

\begin{itemize}
\item{} We perform a quantitative study to investigate the classes and root causes of bugs, which could aid future studies on testing and debugging techniques for Jupyter Notebook projects. 

\item{} We complement this study with a qualitative study on how data scientists professionals perceive the bugs in Jupyter Notebook projects and features missing in the platform.

\item{} A taxonomy that includes eight bug categories for Jupyter Notebook projects. 

\item{} Based on a set of observations from the mining software repository study (GitHub and Stack Overflow) and interviews, we provide some recommendations for researchers, and practitioners.

\item For replication and reproducible research, we make our materials available on our project website. These include a dataset of Jupyter Notebook bugs collected from GitHub and Stack Overflow, and all interview data (prompt, summary of professional and demographic information from Participants, and codebook). Our artifacts can be found at the accompanying website\footnote{https://github.com/bugs-jupyter/empirical-study}.

\end{itemize}

\section{Methodology}

%

This section describes the methodology used in our study to characterize Jupyter Notebook bugs which involve GitHub repository mining, StackOverflow posts analysis, and semi-structured interviews with data scientists. To fulfill this purpose, our study aims to answer the following research questions (RQs):  

\begin{itemize}
    \item RQ1. What types of bugs are more frequent? \textit{Motivation:} It aims to identify the types of bugs and how often they appear. It is the first step toward better understanding and building a taxonomy of bugs in Jupyter Notebooks projects.
    
    \item RQ2. What are the root causes of bugs? 
    \textit{Motivation:} The root cause of bugs provides additional information to understand the bugs better. Comprehending these causes can help understand what is needed to work around, fix, or improve the Jupyter environment. 
    
    \item RQ3. What are the frequent impacts of bugs? 
    \textit{Motivation:} Understanding and quantifying the impact of a bug can help prioritize and scale how severe it is.
    
    \item RQ4. What challenges do data scientists face in practice on Jupyter Projects?
    \textit{Motivation:} The Jupyter Notebook is commonly adopted by data scientists from different domains, from finance systems to the car industry. Despite their growing adoption and popularity, there has been no study to understand Jupyter Notebook usage challenges from practitioners' points of view. The current environmental limitations can also be analyzed.
\end{itemize}

\begin{figure}[H]
\includegraphics[width=8cm]{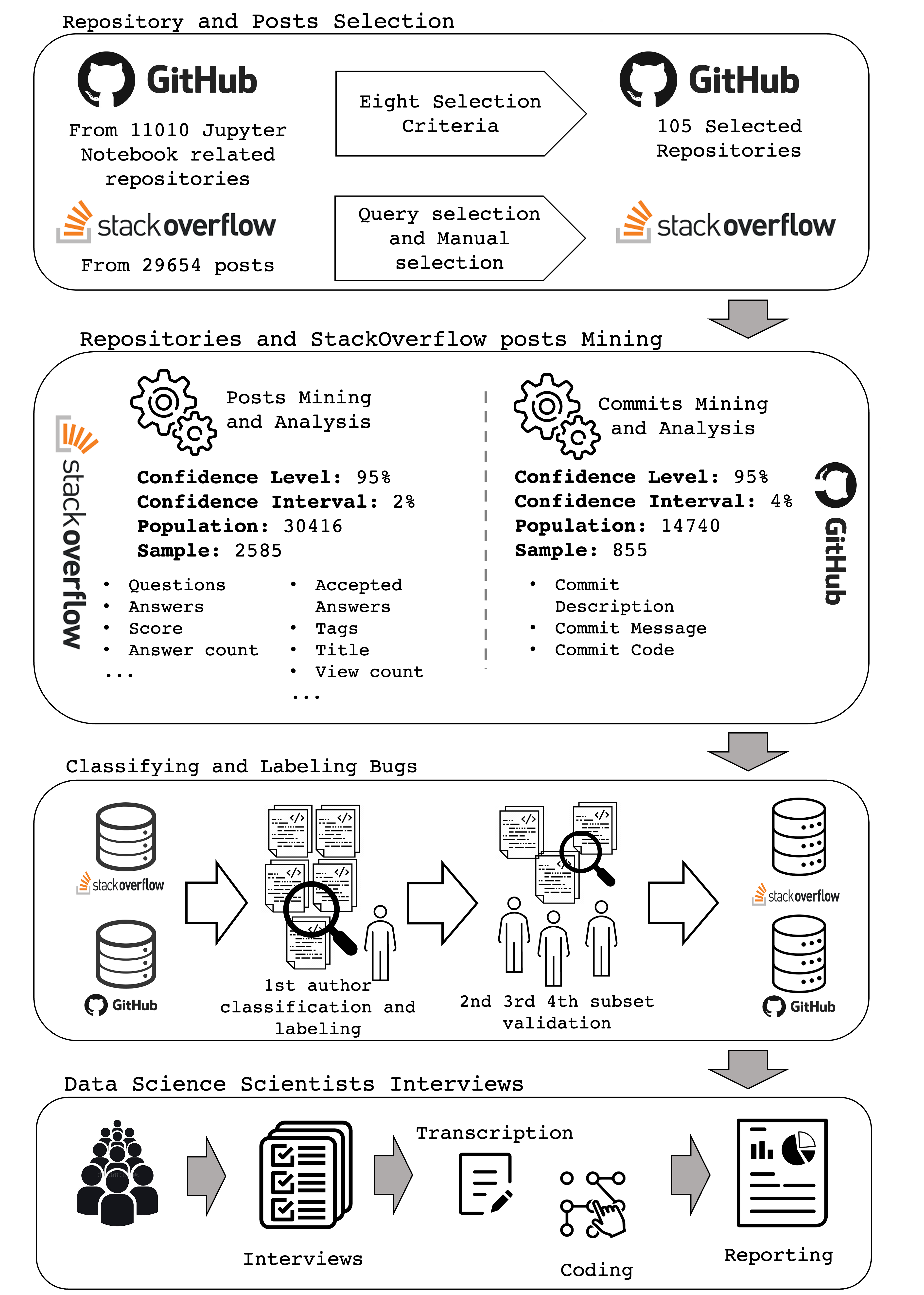}
\caption{Research Methodology.}
\label{methodology}
\end{figure}

In order to answer these questions, the following steps were performed. 
Firstly, a \textit{(i) GitHub\footnote{https://github.com} Repository Mining} analysis was performed to characterize  bugs in the context of Jupyter Notebooks\footnote{https://jupyter.org} projects. In this analysis, only commits related to bug fixing were considered by inspecting the commit message \cite{DBLP:conf/icse/Makhshari021, 10.1145/3338906.3338955, 10.1145/3468264.3468559, DBLP:conf/icse/GarciaF0AXC20}. Next, \textit{(ii) the StackOverflow\footnote{https://en.stackoverflow.com} posts} analysis was performed to characterize data science difficulties/issues/questions when using Jupyter Notebooks. Next,\textit{(iii) manual labeling and classification} were performed to identify the main bug types, root causes, and impact. Finally, \textit{(iv) semi-structured interviews} were conducted with data scientists to obtain and validate insights on the main issues when developing using Jupyter Notebooks projects. Figure \ref{methodology} shows the overall research methodology used in our study. All the quantitative and qualitative data is available online at the accompanying website\footnote{https://github.com/bugs-jupyter/empirical-study}.

\subsection{Repositories and Posts Selection and Mining}

We chose GitHub repositories predominantly written in the "Jupyter Notebook" language in the initial step since we are interested in Data Science projects using this environment. Next, in order to filter out the most relevant and active projects, some inclusion and exclusion criteria were applied as recommended by \cite{DBLP:journals/peerjpre/MunaiahKCN16}.
\begin{itemize}
    \item \textbf{Exclusion Criteria}: 
    \begin{itemize}
        \item The repository must have a description of its main proposal;
        \item Repositories that did not have their artifacts and description in English were not considered in the study; and
        \item Repositories corresponding to tutorials, books, and classroom materials were also removed from our analysis.
    \end{itemize}
\end{itemize}
\begin{itemize}
    \item  \textbf{Inclusion Criteria}: 
    \begin{itemize}
        \item The repository has been updated at least once in 2020; 
        \item The repository must have at least 24 commits in 2020 (corresponding to two commits per month in 2020). This criterion was used to filter out inactive repositories;
        \item The repository must have at least ten contributors in 2020. This criterion was used to eliminate irrelevant repositories, c.f., \cite{DBLP:conf/icse/AgrawalRKSM18}, \cite{DBLP:conf/icse/KrishnaARSM18}, \cite{DBLP:conf/sigsoft/RahmanAKS18}; and
        \item Finally, the repository must have commits in \texttt{.iphnb} files with the following keywords in the commit message: \textit{ 'fix', 'fixes', 'fixed', 'fixing', 'defect', 'defects', 'error', 'errors', 'bug', 'bug fix', 'bugfixing' 'bugfix','bugs', 'issue', 'issues', 'mistake', 'mistakes', 'mistaken', 'incorrect', 'fault', 'faults','flaws', 'flaw', 'failure', 'correction', 'corrections'}. This criteria was used to filtered out commits related to bug fixing. \cite{DBLP:conf/icse/Makhshari021,DBLP:conf/icse/GarciaF0AXC20}. 
    \end{itemize}
\end{itemize}

After filtering, we selected the top 105 Jupyter repositories, resulting in 14740 valid commits in our GitHub raw dataset. Next, the StackOverflow posts were retrieved using the query\footnote{("select Id, PostTypeId, AcceptedAnswerId, ParentId, CreationDate, DeletionDate, Score, ViewCount, OwnerUserId, OwnerDisplayName, LastEditorUserId, LastEditorDisplayName, LastEditDate, LastActivityDate, Tags, AnswerCount, CommentCount, FavoriteCount, ClosedDate, CommunityOwnedDate, ContentLicense, Title from Posts where Tags LIKE jupyter-notebook)} applied to the StackOverflow API resulting in 30416 posts in our StackOverflow raw dataset.

\subsection{Classifying and Labeling Bugs}


We created a spreadsheet with all GitHub commits and StackOverflow posts filtered, containing the bug type, root cause, and impact. The bug type refers to errors found in Jupyter Notebook projects and grouped into categories. The grouping process was iteratively performed by classifying and validating the results. In order to analyze them, we investigated the title, body, pull requests, and other information that can assist us with gaining a comprehensive understanding of issues on GitHub commits. Regarding StackOverflow, we analyzed the title, body, the comments of the selected posts, and also the accepted answers \cite{DBLP:conf/icse/Makhshari021, 10.1145/3377811.3380409, 10.1145/3338906.3338955, DBLP:conf/issre/ThungWLJ12}. 
To explore the root cause, we analyzed the reason that triggered the error by analyzing the changes made in the bug fixing commits, and the answers that provide a solution in the StackOverflow \cite{DBLP:conf/icse/Makhshari021, DBLP:conf/kbse/YangZGK21, DBLP:conf/icse/GarciaF0AXC20, 10.1145/3213846.3213866}. 
Finally, regarding impact, we analyzed major effects of bugs by reading the commit message, pull request messages and the associated issues. In the ScakOverflow, the question description was important to understand the impact \cite{DBLP:conf/icse/GarciaF0AXC20, 10.1145/3338906.3338955, DBLP:conf/issre/ThungWLJ12}.

Once the 14740 commits and 30416 posts were collected, the first and second authors started the open coding process \cite{DBLP:conf/icse/Makhshari021} to evaluate and label the dataset. It was performed until the categories (bug type, root cause, and impact) reached a saturation state where no new categories appeared \cite{saturation}. This saturation was achieved when analyzing 855 of 14740 commits, giving us a margin of error of less than 4\% at 95\% confidence level. In addition, analyzing 2585 of 30416 posts gave us a margin of error of less than 2\% at a 95\% confidence level. Finally, having established the reliability of judgment, new commits and posts were classified by a single author. This reliability and saturation of judgment was achieved with 855 commits and 2585 posts. During this step, some commits were discarded since they were not related to bugs or reported "typo" errors and improvements not related to bug fixing. Some StackOverflow posts were also discarded since they mentioned some hacking or only questions related to Jupyter Notebook usage. Next, three authors (2nd, 3rd and 4th) independently classified 145 commits randomly, and 137 posts were selected to validate the first author classification. We measured the inter rater agreement among the authors using Cohen’s Kappa coefficient \cite{6676898}. A training session was performed among the authors to clarify the labeling and what they mean. After that, the Cohen’s Kappa coefficient was more than 81\% for bug type, 95\% for root cause and 95\% for impact, which according to Landis and Koch \cite{Landis77}, is ‘substantial agreement’.


\subsection{Data Scientists Interviews}

To validate the findings identified in the previous steps and understand how these bugs impact the daily life of data scientists working with Jupyter notebook projects, we conducted semi-structured interviews with data scientists. 


\textbf{Protocol.} We designed the interview prompt to understand and validate previous findings related to the data scientists' usage of Jupyter notebook projects. It was composed of eighteen open questions. The participants were informed that they could omit to answer a question to avoid arbitrary answers. The interviews started with some demographic questions and participants' expertise. The technical section is composed of questions related to the Jupyter Notebook environment and the tool's problems and challenges.

The interview pilot was performed using one data scientist. After that, the 2nd and 3rd authors also support the interview improvement, solving questions difficulties based on pilot feedback. Some questions were added, updated, and removed to make the interview easier to understand and answer. The pilot interview responses were only used to calibrate the instrument, and these responses were not included in the final results. The interview instrument can be seen in the supplementary material attached to the paper. All the interviews were conducted remotely, and we recorded the audio to further analysis with the participants' consent. The interviews took about 43 minutes on average. We transcribed the recorded interviews using QDA Miner\footnote{https://provalisresearch.com/products/qualitative-data-analysis-software/freeware/}. 

\textbf{Participants.} After conducting a pilot interview with one data scientist (not included in the study) as a pretest \cite{Seidman2006}, nineteen data scientists were interviewed. All of them have at least one year of data science experience from several companies and domains as seen in Table \ref{inter}. The data scientists came from 12 different companies, working in domains such as mobile games, finance, car, petrochemical, mining, etc. 40\% of participants hold a Ph.D., 25\% holds a master's degree, 25\% hold a bachelor's degree, and 10\% conducted post doctoral studies.

\begin{table}[ht!]
\footnotesize
\begin{tabular}{ c c c c c }

\hline
 \textbf{Id} & \textbf{Role} & \textbf{Company Area} & \textbf{Exp. (Years)} \\ 
 \hline
 DS1 & Data Engineer & Petrochemical Industry & 5 \\ 
 DS2 & Feature Owner & Car Industry & 8 \\ 
 DS3 & Data Scientist & Finance & 13 \\ 
 DS4 & Coordinator & Mining Company & 10 \\ 
 DS5 & Data Scientist & Finance & 8 \\ 
 DS6 & Software Engineer & Engineering solutions & 10 \\
 DS7 & Data Scientist & Mobile Games & 11 \\
 DS8 & IA Researcher & University & 20 \\
 DS9 & Data Scientist & IT Services & 18 \\
 DS10 & Teacher& University & 15 \\
 DS11 & ML Engineer & Mobile Games & 13 \\
 DS12 & Data Scientist & Finance & 12 \\
 DS13 & Data Scientist & Finance & 25 \\
 DS14 & Business Manager & Finance & 17 \\
 DS15 & Data Scientist & Finance & 15 \\
 DS16 & Data Scientist & Finance & 14 \\
 DS17 & Data Scientist & Finance & 11 \\
 DS18 & Data Scientist & Finance & 9 \\
 DS19 & DS Researcher & University & 9 \\
 \hline 

\end{tabular}
\caption{\label{inter}Interview Participants background.}
\end{table}

\textbf{Analysis.} The audio transcription was the first step (14 hours and 33 minutes). The first author was responsible for conducting the transcription process using the OTranscribe \footnote{https://otranscribe.com/} tool. We also performed a minor review to validate the transcriptions and clarify some answers.

Next, the first author started the coding process using the QDA Miner Lite \footnote{https://provalisresearch.com/products/qualitative-data-analysis-software/freeware/} tool. The first author and two experts iteratively worked in the coding step to reduce the subjective bias during the open coding process. We used a set of first-cycle, and second-cycle coding methods for data analysis \cite{saldana2015coding}. The first cycle methods are those processes during the initial coding of data. Second-cycle methods, if needed, are ways of reorganizing and reanalyzing data coded through first-cycle methods. All codes created in our study were later on clustered into categories. Analyzing our data, we could define categories to understand the answers from interview participants. Next, the authors resolved the potential conflicts in the labels and categories. It resulted in 52 codes, 7 categories, and 5 challenges.





\section{Results}

This section reports the answers to our targeted research questions and findings from the data collected from GitHub, StackOverflow, and interview responses.

\subsection{Types of Bugs in Jupyter Projects (RQ1)}

\begin{center}
\begin{figure}[H]

\includegraphics[width=\columnwidth]{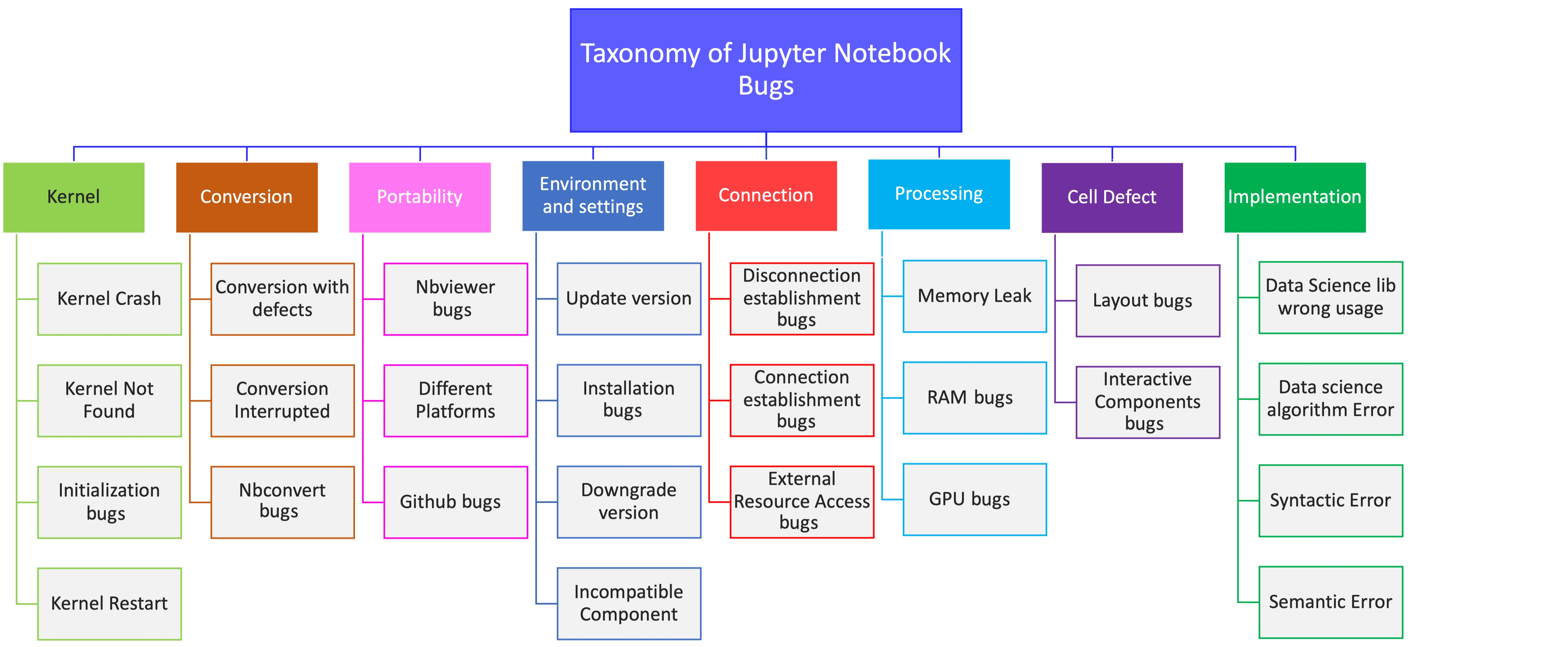}
\caption{Taxonomy of Jupyter Notebook bugs.}
\label{Taxonomy}

\end{figure}

\end{center}

Data scientists face different types of bugs when using Jupyter notebooks. To understand these bugs, we classified them into different types and created an initial taxonomy. Next, we used the interviews to validate and improve the proposed taxonomy. Figure \ref{Taxonomy} shows the taxonomy, and then we describe the types of bugs with examples and their occurrence percentage (in parenthesis) in StackOverflow and GitHub.
\textbf{Kernel Bugs | KN - \textit{(StackOverflow - 10.8\% | GitHub - 2.9\%)}.} 
This type covers the bug problems in the kernel operation when using Jupyter Notebooks. The most common occurrences of Kernel Bugs are crashing, booting, installation, and unresponsive problems.

\begin{itemize}
    \item \textit{Kernel Crash:} A common bug that happens during notebook usage is when the kernel breaks. Sometimes the crashing is followed by a warning message, and the kernel is unusable in other cases. According to participants in our interviewees, it is a common bug fixed by kernel restarting
    \item \textit{Kernel Not Found:} It happens when the user starts the Jupyter Notebook, but it is not linked to a Kernel. This way, the Kernel not found message is displayed. Some StackOverflow posts relate this bug to installation issues.
    \item \textit{Initialization Bugs:} It happens during kernel initialization, usually caused by wrong installations or conflict with the installed kernel.
    \item \textit{Kernel Restart:} The kernel unexpectedly restarts during its usage.
\end{itemize}
   
\underline{\textit{Example: }}Kernel bugs can cause many problems, such as data and information loss and, delays in project time, repeating the lost analyses. In bug \#107937815 (GitHub) or \#35673530 (StackOverflow) where the Python updating generated incompatibility among packages used in the notebook and as reported by DS13:\\

\checkmark \textit{ DS13: " The Kernel bugs are the most frequent ones (...). It impacts project execution time since it interrupts the data analysis."}\\


\textbf{Conversion | CV - \textit{(StackOverflow - 6.7\% | GitHub - 10.6\%)}
.} It comprehends bugs related to errors during notebook conversion from .ipynb file type to other formats. Data scientists commonly use the conversion function to distribute their analyses and results to different audiences. This type refers to bugs during conversion, resulting in poorly rendered conversions or corrupted files.
\begin{itemize}
    \item \textit{Conversion Interrupted:} It occurs when there is an attempt to convert a notebook to another format, but this conversion is interrupted.
    \item  \textit{Conversion with defects:} It happens when a conversion task finishes successfully, but its result contains unintended defects, for example, PDFs generated without images.
    \item \textit{Nbconvert bugs:} It is related to bugs from nbconvert module, responsible for conversions using command line commands. In these cases, the conversation does not even start.
\end{itemize}
 
\underline{\textit{Example: }} Conversion is one of the essential Jupyter functionalities. Bugs \#99244384 (GitHub) and \#46415269 (StackOverflow) are examples involving the nbconvert module. It impacts
the user experience, mainly for new users, as reported by DS12:\\ 

\checkmark \textit{ DS12: "It happens with new users, which spend considerable time performing the export procedure."}\\

\textbf{Portability | PB - \textit{(StackOverflow - 2.7\% | GitHub - 1.3\%)}.} It involves bugs that are related to Jupyter notebook execution in different environments. Although this feature is one of the pillars in the Jupyter project \cite{ProjectJupyter}, we found different bug occurrences, such as compatibility, rendering, and environment configuration problems. Thus, this bug refers to errors obtained when rendering the notebook in environments and platforms other than the original one. 


\begin{itemize}
    \item \textit{GitHub Bugs.} It is related to bugs when executing .ipynb files in the GitHub environment. It happens when the notebook is not rendered or shows rendering defects.
    \item  \textit{Nbviewer Bugs.} Similar to the previous one, it happens when the user tries to execute the .ipynb file in the nbviewer platform. 
     \item  \textit{Different Platforms:} This bug is related to the attempt to run the notebook on a different platform from its origin, which can happen in situations of different Operating Systems, machines, and browsers (even situations of execution of a .ipynb in a Google Colab, Jupyter-Lab or any platform other than the original). This bug is generally related to the difference in configurations between the platform it was originally developed on and the platform it was ported to.
\end{itemize}

\underline{\textit{Example: }} Problems at this stage make it difficult to disseminate the analysis. Bugs \#200722670 (GitHub) and \#47868625 (StackOverflow) describe the need for modifications to correctly display the notebook on the GitHub environment. Participant DS11 reported a similar problem:\\

\checkmark \textit{ DS11: "GitHub has a tool to view Jupyter notebooks, right, but it's kind of random, it opens whenever it wants. It doesn't always work to open Jupyter notebook in the browser."}\\

\textbf{Environments and Settings | ES - \textit{(StackOverflow - 43.2\% | GitHub - 35.6\%)}.} It is related to bugs in the development environment and configuration issues. It can happen due to several aspects, such as missing libraries, issues during libraries installation, deprecated libraries, incompatibility between components and libraries, incompatibility with operational systems, problems with a package manager(such as Anaconda, PIP), and problems with installation and configuration of extensions.

\begin{itemize}
    \item \textit{Update and Downgrade Version:} It happens due to incompatibility with the currently installed version of a library or extension, and this library or extension needs to be updated or downgraded to work correctly.
    \item \textit{Installation Bugs:} Wrong installations may cause this bug or lack of dependencies during installation.
    \item \textit{Incompatible Component:} The components used in notebooks can be different, and some of them or versions of some of them generate incompatibilities for use in the same notebook. When installed or used, reports of extensions generate incompatibilities with various components.
\end{itemize}

\underline{\textit{Example: }} The environment setup is a time-consuming task. The Bug \#200722670 (GitHub) and \#35561126 (StackOverflow) show the cost of solving a problem due to a wrong Python version. Participant DS13 suggested that the notebook could aid the user with this setup checking:\\

\checkmark \textit{ DS13: "Depending on the project you're working on and the dependencies you need to install, the setup environment is a laborious task. Maybe it could be managed by Jupyter Notebooks. It is hard to say how it would be possible, but the environment creation could help to avoid configuration problems and everything else."}\\

\textbf{Connection Bugs | CN - \textit{(StackOverflow - 6.2\% | GitHub - 0.9\%)}.} It happens when connecting the notebook with external resources, such as databases, hardware, and repositories. It can occur in two ways:

\begin{itemize}
    \item \textit{External Resource Access Bugs}: It happens when the notebook disconnects or is no longer available to external resources.
    
    \item \textit{Disconnection and Connection Establishment Bugs}: In this bug, the notebook itself loses connection to its server.
\end{itemize}

\underline{\textit{Example: }} The Bugs \#107937815 ( GitHub) and \#63863571 (StackOverflow) report problems related to url and external image connection. Another connection problem reported happens when receiving data through a serial port, as highlighted by Participant DS2:\\

\checkmark \textit{ DS2: "... during Arduino usage some problems are difficult to know the root cause. It this situation, we looked at the Arduino board, try to disconnect and connect again, turn it on and off, replace the Arduino board to see if one of the work around solve our problem. After all tries, for some reason we get the Arduino board connected to Jupyter notebook."}\\

\textbf{Processing | PC - \textit{(StackOverflow - 4.9\% | GitHub - 1.9\%)}}. Data analysis often requires high processing power. Thus, memory availability and concurrency are valuable resources. Bugs of this type are related to Timeout, Memory Errors, and longer processing tasks.  
 
\begin{itemize}
    \item \textit{Memory Leak}: It occurs when a large memory allocation is incompatible with the process that is being executed. In general, the user identifies this bug when there is a delay in the execution. 
    \item \textit{RAM and GPU Bugs}: All bugs related to memory overflow and slow processing fall into this category.
\end{itemize}

\underline{\textit{Example: }} This bug may affect data scientists by increasing time analysis, interruptions, and data loss. Bugs \#86884600 (GitHub) and \#643288550 (StackOverflow) report a bug related to high-resolution images, in which a workaround should be performed to get the notebook processed. Chattopadhyay et. al \cite{DBLP:conf/chi/ChattopadhyayPH20} also reported Jupyter lack of support for handling large volumes of data, and one of our participants also reported this:\\

\checkmark \textit{ DS10: "It has happened several times with me, and it happened when I was manipulating large datasets. I spent some time understanding, debugging, and identifying the root cause of this bug."}\\

\textbf{Cell Defect (CD) - \textit{(StackOverflow - 3.6\% | GitHub - 2.6\%)}}. It involves bugs related to notebook cell rendering, such as code cells, markdown, or outputs, and it usually happens when using interactive components, latex, markdown, or cells. Next, we present some groups of this bug.

\begin{itemize}
    \item  \textit{Layout Bugs}: It refers to cell rendering problems, such as results beyond the margin, unexpected formulas, testing formatting, blank cells, graphics visualization problems, and so on. It can happen in any Jupyter notebook cell.  

    \item \textit{Interactive Components Bugs}: It happens with components that allow the users to interact directly with the rendered cells. 
\end{itemize}

\underline{\textit{Example: }} Bugs \#237890763 (GitHub) and \#69695030 (StackOverflow) are examples in which the user faces problems with "input()" or notebook scrollbar. Participant DS14 highlighted this as follows:\\

\checkmark \textit{ DS14: "It was a very annoying error, and it frequently occurs on a personal computer as a Mac. For some reason, the cell size reduced and ended up cutting the text in half. I don't know, I could not identify what caused it (...) and it happens a lot."}\\

\textbf{Implementation | IP - \textit{(StackOverflow - 22\% | GitHub - 44.2\%)}}. Bugs related to implementation in general, syntax, logical, non-instantiated variables, algorithms, and semantics are examples of this type of bug. Analyzing all the posts and commits, we identified the following implementation bugs:

\begin{itemize}
    \item \textit{Semantic Error}: Bugs related to logic misunderstandings. In this bug, the code executes correctly, but its execution generates a different output than expected, either due to poorly defined parameters or wrong algorithms.
    
    \item  \textit{Syntax Error}: Programming bugs include incorrect variable or function declaration and calls, missing or incorrectly assigned parameters, missing or misplaced parentheses, warnings or errors generated by nonstandard Python (PEP8) coding, and other general programming errors.
    
    \item \textit{Data Science lib wrong usage}: Bugs related to the inappropriate use of functions from typical data science libraries, such as Pandas, Scikit-learn, TensorFlow, and so on.
    
    \item \textit{Data Science Algorithm Error}: Bugs in the logic of statistical analysis or machine learning models.
\end{itemize}

\underline{\textit{Example: }} The implementation bugs are common for developers, but in Jupyter notebooks, it is potentialized by the possibility of creating duplicated cells and cells out of order. Bugs \#222507066 (GitHub) and \#45946060 (StackOverflow) are examples where changes were made to fix errors of duplicate code and out-of-order cell execution. Participant DS14 reported this bug as follows:\\

\checkmark \textit{ DS14: "When you're writing in your notebook, you can write your code along with your text and it's easy to lose context at some point. For example, if you write in a cell at the top of the notebook, keeping the context of the cells running part bottom of the notebook, when you run your code nothing will make sense."}\\

\begin{figure}[H]\includegraphics[width=8cm]{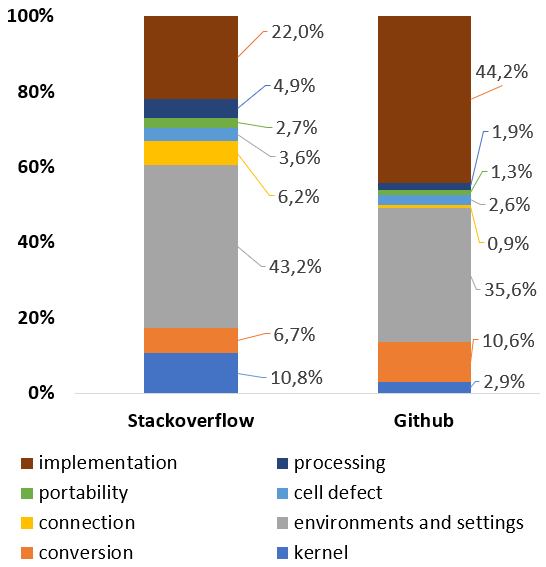}
\caption{Frequency of bug types}
\label{frequence_bugtype}
\end{figure}

\textbf{Frequent bug types -} In order to understand the frequency of each bug type previously discussed, we statistically analyzed the labeled data. Figure \ref{frequence_bugtype} shows the distribution of bug types in GitHub and StackOverflow. 

The most frequent bug in both datasets (GitHub and StackOverflow) was the "Environment and Settings" with 35.6\% and 43.2\%, respectively. It was also reinforced by the interviewers, which highlighted problems with version control, component incompatibility, wrong or missing installations, and problems with extensions. The second most frequent bug type was "Implementation" with 44.2\% (GitHub) and 22\% (StackOverflow). 



We calculated the average annual growth (2014 - 2021) in the StackOverflow dataset for a more in-depth analysis of the bug types and their occurrences. We calculate the annual average growth by first calculating the annual growth per year, then calculating the annual average growth. The following formulas were used to compute this growth. 
\begin{quote}
$
GR\textsubscript{15-14} = (\frac{FinalValue\textsubscript{2015} - StartValue\textsubscript{2014}}{StartValue\textsubscript{2014}})*100
$
\end{quote}

Where FinalValue2015 is the number of occurrences at the end of 2015 and StartValue2014 is the number of occurrences at the beginning of 2014. The annual average growth was calculated, as shown in the formula below, by calculating the average growth considering all ranges.

\begin{quote}
$
AGR = \frac{GR\textsubscript{15-14} + GR\textsubscript{16-15} + GR\textsubscript{17-16} +
GR\textsubscript{18-17} +
GR\textsubscript{19-18} +
GR\textsubscript{20-19} +
GR\textsubscript{21-20}}{7}
$
\end{quote}

\begin{figure}[H]
\includegraphics[width=12cm]{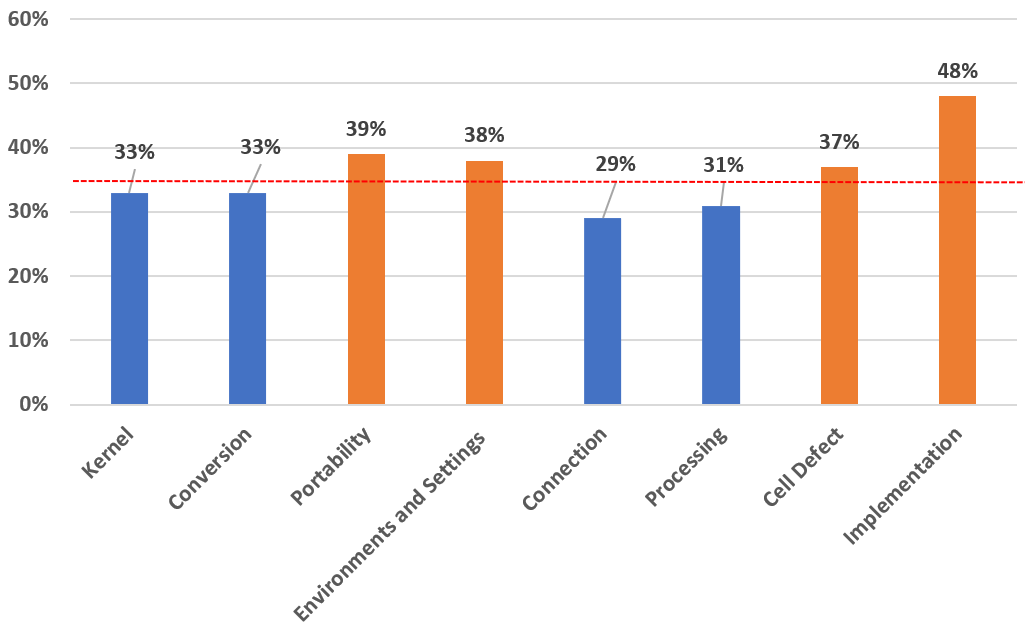}
\caption{Average annual growth of bugs.}
\label{avg_anual_bug}
\end{figure}

Four types of bugs are growing above the general annual average (see Fig. \ref{avg_anual_bug}).

The "Implementation" and "Environment and Settings" bugs grow at a rate of 48\% and 38\%, respectively, which is reflected in the total percentage of the number of occurrences. The two bugs correspond to more than 60\% of the total bug occurrences in the two analyzed databases (see Fig. \ref{frequence_bugtype}).

However, the "Portability" and "Cell Defect" bug types show an annual growth rate higher than the total average, 39\% and 37\%, respectively, despite a low overall occurrence rate.


Previous research \cite{DBLP:conf/msr/PimentelMBF19,PimentelMBF21, DBLP:conf/icse/WangLZ21} analyzed exceptions related to reproducibility errors to deepen the view of the errors found. Likewise, as many of the bug reports analyzed in this study have exceptions encountered by users, we correlate these exceptions with the types of bugs reported to understand them better.

The exceptions reported were also collected and analyzed to understand better the most frequent type of bugs (see Table \ref{exceptions}). While the "ImportError", "ModuleNotFoundError" and "AtributeError" frequently occur in Environment and Settings bugs, the "TypeError", "AttributeError" and "NameError" appears on Implementation bugs. Table \ref{exceptions} summarizes the main problems found in each type of bug.



\begin{table}[H]
\begin{tabular}{ p{3cm} p{0.5cm} p{0.5cm} p{0.5cm} p{0.5cm} p{0.5cm} p{0.5cm} p{0.5cm} p{0.5cm} }
\hline
 \textbf{Exception} & \textbf{ES} & \textbf{IP} & \textbf{KN} & \textbf{CN} & \textbf{PC} & \textbf{CV} & \textbf{PB} & \textbf{CD} \\ 
 \hline
ImportError & 297 & 3 & 11 & 8 & 0 & 4 & 0 & 0\\
ModuleNotFoundError & 266 & 5 & 5 & 3 & 0 & 0 & 2 & 0\\
AttributeError & 116 & 101 & 2 & 8 & 1 & 1 & 0 & 1\\
TypeError & 31 & 222 & 4 & 0 & 0 & 1 & 0 & 0\\
OSError & 17 & 8 & 2 & 2 & 1 & 3 & 0 & 0\\
RuntimeError & 16 & 2 & 2 & 0 & 7 & 1 & 0 & 0\\
NameError & 14 & 76 & 1 & 0 & 0 & 0 & 0 & 0\\
FileNotFoundError & 14 & 9 & 12 & 2 & 0 & 2 & 0 & 0\\
ValueError & 12 & 15 & 3 & 1 & 0 & 4 & 0 & 0\\
SyntaxError & 2 & 1 & 0 & 1 & 0 & 0 & 1 & 0\\
 \hline 
\end{tabular}
\caption{\label{exceptions}Python Exceptions per Type of Bugs.}
\end{table}

\begin{mybox}{Finding 1}
The most frequent bugs in the Jupyter notebook are those related to Environments and Settings (StackOverflow - 43.2\% | GitHub - 35.6\%) and Implementation (StackOverflow - 22\% | GitHub - 44.2\%), which also show an annual growth rate above the average, 38\% and 48\% respectively. Although the Portability and Cell Defect bugs have had fewer occurrences, they have had above-average growth over the years.
\end{mybox}

\subsection{Root Causes of Bugs (RQ2)}

The root cause of bugs helps us understand their origin and how we can correct them. Table \ref{frequence_rootcause} shows the distribution of bug types according to their root causes. As can be seen in the table, the three more frequent root causes are Install and Configuration Problems (StackOverflow - 32.1\% | GitHub - 16.3\%), Version Problems (StackOverflow - 19.0\% | GitHub - 22.5\%) and Coding error (StackOverflow - 17.6\% | GitHub - 31.5\%). These are bugs whose causes are related to component installation or configuration problems, wrong component versions, and coding problems such as semantic, logical, or syntax errors. 


\begin{table}[H]
\footnotesize
\begin{tabular}{ p{2cm}cccccccccccccccc }
\hline
 \textbf{Root Causes} & \multicolumn{2}{|c|}{\textbf{KN}} & \multicolumn{2}{|c|}{\textbf{CV}} & \multicolumn{2}{|c|}{\textbf{ES}} & \multicolumn{2}{|c|}{\textbf{CN}} & \multicolumn{2}{|c|}{\textbf{CD}} & \multicolumn{2}{|c|}{\textbf{PB}} & \multicolumn{2}{|c|}{\textbf{PC}} & \multicolumn{2}{|c|}{\textbf{IP}} \\ 
 \hline
 & SO&GI&SO&GI&SO&GI&SO&GI&SO&GI&SO&GI&SO&GI&SO&GI\\
 \hline
Install and Configuration Problems & \multirow{ 3}{*}{56} & \multirow{ 3}{*}{12} & \multirow{ 3}{*}{44} & \multirow{ 3}{*}{0} & \multirow{ 3}{*}{635} & \multirow{ 3}{*}{117} & \multirow{ 3}{*}{42} & \multirow{ 3}{*}{0} & \multirow{ 3}{*}{6} & \multirow{ 3}{*}{7} & \multirow{ 3}{*}{12} & \multirow{ 3}{*}{3} & \multirow{ 3}{*}{0} & \multirow{ 3}{*}{0} & \multirow{ 3}{*}{36} & \multirow{ 3}{*}{0} \\
\hline
Version Problems & 45 & 7 & 24 & 2 & 374 & 158 & 30 & 0 & 11 & 0 & 6 & 0 & 0 & 0 & 0 & 25 \\ 
\hline
Deprecation & 1 & 0 & 0 & 0 & 21 & 1 & 1 & 0 & 0 & 0 & 0 & 0 & 0 & 0 & 1 & 0 \\ 
\hline
Permission denied & \multirow{ 2}{*}{2} & \multirow{ 2}{*}{0} & \multirow{ 2}{*}{1} & \multirow{ 2}{*}{0} & \multirow{ 2}{*}{16} & \multirow{ 2}{*}{0} & \multirow{ 2}{*}{1} & \multirow{ 2}{*}{1} & \multirow{ 2}{*}{0} & \multirow{ 2}{*}{0} & \multirow{ 2}{*}{0} & \multirow{ 2}{*}{0} & \multirow{ 2}{*}{1} & \multirow{ 2}{*}{0} & \multirow{ 2}{*}{3} & \multirow{ 2}{*}{0} \\ 
\hline
TimeOut & 5 & 0 & 9 & 0 & 4 & 0 & 18 & 0 & 0 & 0 & 0 & 0 & 6 & 1 & 7 & 0 \\ 
\hline
Memory Error & 34 & 0 & 2 & 0 & 3 & 0 & 7 & 0 & 0 & 1 & 1 & 0 & 78 & 12 & 20 & 0 \\ 
\hline
Coding error & 4 & 3 & 12 & 49 & 17 & 18 & 4 & 4 & 17 & 7 & 4 & 0 & 0 & 3 & 397 & 185 \\ 
\hline
Logic error & 9 & 1 & 0 & 0 & 2 & 1 & 0 & 0 & 4 & 2 & 0 & 0 & 4 & 0 & 32 & 114 \\ 
\hline
Hardware software limitations & \multirow{ 2}{*}{5} & \multirow{ 2}{*}{0} & \multirow{ 2}{*}{20} & \multirow{ 2}{*}{39} & \multirow{ 2}{*}{12} & \multirow{ 2}{*}{0} & \multirow{ 2}{*}{19} & \multirow{ 2}{*}{1} & \multirow{ 2}{*}{33} & \multirow{ 2}{*}{4} & \multirow{ 2}{*}{32} & \multirow{ 2}{*}{8} & \multirow{ 2}{*}{36} & \multirow{ 2}{*}{0} & \multirow{ 2}{*}{15} & \multirow{ 2}{*}{0} \\ 
\hline
Unknown & 117 & 2 & 60 & 1 & 34 & 9 & 38 & 2 & 22 & 1 & 14 & 0 & 1 & 0 & 58 & 54 \\ 
 \hline 
\end{tabular}
\begin{minipage}{15cm}
\footnotesize{(KN) Kernel, (CV) Conversion, (PB)  Portability, (ES) Environment and Settings, (CN)  Connection, (PC) Processing, \\(CD) Cell Defect, (IP)  Implementation.}
\end{minipage}
\caption{\label{frequence_rootcause}Frequency of Bug Type vs Root Cause}

\end{table}

\begin{table}[H]
\footnotesize
\begin{tabular}{ p{2cm}cccccccccccccccc }
\hline
 \textbf{Impacts} & \multicolumn{2}{|c|}{\textbf{KN}} & \multicolumn{2}{|c|}{\textbf{CV}} & \multicolumn{2}{|c|}{\textbf{ES}} & \multicolumn{2}{|c|}{\textbf{CN}} & \multicolumn{2}{|c|}{\textbf{CD}} & \multicolumn{2}{|c|}{\textbf{PB}} & \multicolumn{2}{|c|}{\textbf{PC}} & \multicolumn{2}{|c|}{\textbf{IP}} \\ 
 \hline
 & SO&GI&SO&GI&SO&GI&SO&GI&SO&GI&SO&GI&SO&GI&SO&GI\\
 \hline
Crash & 275 & 11 & 49 & 0 & 142 & 16 & 142 & 0 & 4 & 0 & 8 & 0 & 8 & 1 & 1 & 0 \\ 
\hline
Bad Performance & 1 & 0 & 2 & 0 & 7 & 1 & 2 & 0 & 1 & 3 & 0 & 0 & 46 & 14 & 18 & 46 \\ 
\hline
Incorrect Functionality & \multirow{ 2}{*}{0} & \multirow{ 2}{*}{10} & \multirow{ 2}{*}{96} & \multirow{ 2}{*}{91} & \multirow{ 2}{*}{41} & \multirow{ 2}{*}{90} & \multirow{ 2}{*}{7} & \multirow{ 2}{*}{8} & \multirow{ 2}{*}{86} & \multirow{ 2}{*}{19} & \multirow{ 2}{*}{55} & \multirow{ 2}{*}{11} & \multirow{ 2}{*}{1} & \multirow{ 2}{*}{0} & \multirow{ 2}{*}{64} & \multirow{ 2}{*}{261} \\ 
\hline
Run Time Error & 2 & 4 & 24 & 0 & 900 & 197 & 9 & 0 & 2 & 0 & 6 & 0 & 71 & 1 & 472 & 65 \\ 
\hline
Warning & 0 & 0 & 1 & 0 & 28 & 0 & 0 & 0 & 0 & 0 & 0 & 0 & 0 & 0 & 14 & 6 \\ 
 \hline 
\end{tabular}
\begin{minipage}{15cm}
\footnotesize{(KN) Kernel, (CV) Conversion, (PB)  Portability, (ES) Environment and Settings, (CN)  Connection, (PC) Processing, \\(CD) Cell Defect, (IP)  Implementation}
\caption{\label{frequence_impact}Frequency of Impact vs Root Cause}
\end{minipage}

\end{table}

We could not identify all the root causes for some bugs in our dataset. Thus, some bugs were classified using the Unknown category (StackOverflow - 13.3\% | GitHub - 8.1\%). This category is present in all bug types, especially those related to Kernel. It can reinforce the user's difficulty in understanding that bug type.


The root cause of Hardware and Software Limitations (StackOverflow - 6.7\% | GitHub - 6.1\%) occurs when there are limitations in the software or hardware where the notebook is running. It happens in all types of bugs; however, the Memory Error (StackOverflow - 5.6\% | GitHub - 1.5\%) frequently appears as a root cause of the processing bugs.

The other root causes occur occasionally, such as Logic error (2\%) in the developed code; TimeOut (StackOverflow - 1.9\% | GitHub - 0.1\%), when an active process achieves time limit; Deprecation (StackOverflow - 0.9\% | GitHub - 0.1\%), where a component or functionality is outdated; and Permission denied (StackOverflow - 0.9\% | GitHub - 0.1\%), when the permission to access an external resource is denied.

\begin{mybox}{Finding 2}
The most frequent Root Causes in Jupyter Notebook projects are: Configuration issues (StackOverflow - 32.1\% | GitHub - 16.3\%), Version issues (StackOverflow - 19.0\% | GitHub - 22.5\%) and Coding Error (StackOverflow - 17.6\%) | GitHub - 31.5\%) they are the cause of most Implementation and Environments and Settings bugs. The root cause Unknown (StackOverflow - 13.3\% | GitHub - 8.1\%) appears more related to Kernel bugs suggesting a difficulty in identifying its cause.

\end{mybox}

\subsection{Impacts of Bugs (RQ3)}
The impact caused by a bug can help increase its severity and serve as a prioritization model and alert for users. Table \ref{frequence_impact} shows the distribution of bug types according to their impact. The most frequent impacts are Run Time Errors, Incorrect Functionality, and Crashes.

The Run Time Error was the most frequent impact in the StackOverflow (57.5\%) dataset and the second most frequent in GitHub (31.2\%). It is characterized by execution failures followed by an error message. It is commonly found in the Environments and Settings and Implementations types of bugs. 


The impact Incorrect Functionality, which is characterized by bugs that the code can be executed, but the result is not what was expected, had the highest occurrence on GitHub (57.3\%) and appeared on StackOverflow as the third largest impact (13.5\%).

The Crash, as mentioned before, it happens when an interruption in the normal operation or startup of the notebook occurs without any error message, exception, or warning. Considering the GitHub dataset, it happens (3.3\%) in a smaller amount than on StackOverflow (24.3\%). A possible explanation for this is that Crash is an impact that happens more with Kernel bugs, and one of the solutions to solve Kernel Crashes is restarting Kernel, which is not showing up in fix commits. 

Kernel Crash was the only bug/impact mentioned by all participants in our interview session. According to users, even being an annoying bug, it is easy to get around it just restarting the kernel:\\

\checkmark \textit{ DS12: “ The Kernel Crash happens a lot, especially when the memory runs out and the notebook crashes, we need to run it all over again."\\}

\checkmark \textit{ DS7: "The Kernel Crash, is usually solved by restarting and returning back to work and that's ok..."\\}

The other impacts had a smaller volume of occurrences. Bad Performance bugs (StackOverflow - 3.0\% | GitHub - 7.5\%), whose occurrence does not prevent the correct execution, but decreases the quality or performance and Warning (StackOverflow - 1.7\% | GitHub - 0.7\%), which does not impact on notebook functioning, but triggering an alert for the user.

\begin{mybox}{Finding 3}
The most frequent impacts from bugs in Jupyter notebooks are: Run Time Error (StackOverflow - 57.5\% | GitHub - 31.2\%), Incorrect Functionality (StackOverflow - 13.5\% | GitHub - 57.3\%), and Crash (StackOverflow - 24.3\% | GitHub - 3.3\%). They are the effects related to bug types Environments and settings, Implementation and Kernel bugs. The Kernel Crash is a common bug/impact in the daily activities of Jupyter users and has as the main workaround solution, the restart of the Kernel.

\end{mybox}

\subsection{Challenges in Jupyter Notebook Projects (RQ4)}

Previous sections described the bugs types and their frequency. Based on that, a taxonomy was proposed in the Jupyter Notebook domain. Next, we evaluated the root causes and impact and how they affect each bug type. This analysis gives us valuable insights into how bugs are distributed. Finally, to better understand how it affects data scientists' daily lives, we conducted interviews with industry professionals who use Jupyter notebooks in industrial projects.
 
Data science is a multidisciplinary area involving physicists, mathematicians, statisticians, IT professionals, and others. This diversity is also observed in computational notebooks usage. Thus, we interviewed professional Jupyter users from the industry to understand the dynamics of bugs in Jupyter Notebook projects. We used the interviews to validate our results from mining and collect insights, impressions, and challenges about environmental usage. Next, the main challenges identified by professionals are discussed.

\textbf{Backgrounds, and requirements.} How users realize the bug can influence how they fix it. Kim et al. \cite{10.1145/2884781.2884783} highlighted the existence of a diversity of profiles of data science professionals. This diversity makes it possible for many Jupyter users not to come from the computing field or even not have enough experience to feel the need to follow patterns and strategies that help to reduce or identify errors.

Users with less experience or knowledge tend to produce messy, dirty notebooks that can eventually generate errors, having the challenge of using a tool with a simplistic layout (compared to a traditional IDE), without forgetting development standards that bring gain in code quality and consequently reduction of errors. 

Analyzing the interview responses with demographic data (Table \ref{inter}), we realized that, for example, software engineering knowledge is important for identifying the root cause and fixing the bug as highlighted by a professional:\\ 

\checkmark \textit{DS3: "Another very common thing is the knowledge that the person has. Data science is kind of a combination of statistics and computing and within that world you see people from physics, engineering and so on. The concern with having a structured, readable, documented code usually comes from the computing area, as the guy studied software engineering. So you take these people, they have an organized code."}\\



\textbf{Software Quality. } Due to user diversity, some of them lack software engineering practices. Jupyter notebooks increase it since the environment allows users to duplicate cells, drag and drop cells to different locations, and so on. In addition, notebooks do not provide any mechanism to control and support the users in this respect. Although Jupyter provides flexibility and allows users with different backgrounds to use it, no support is provided regarding code quality. This aspect was also mentioned by a professional:\\

\checkmark \textit{DS14: "The lack of some functionality can be a problem, it can discourage the data scientist writing better code, using good software engineering practices. I see this a lot, my codes when I'm writing in VSCode, for example, are much better than when I'm writing in Jupyter, I feel this also happens in RStudio (...) I can write code better in an IDE than in the Jupyter."}\\

\textbf{Testing and debugging. } Software debugging is an essential activity to improve code quality. Some interviewees pointed out the lack of basic debugging and testing tools as a challenge to be addressed. They also detailed the process of fixing a bug using a trial and error approach. Among the interviewees, especially those with a software engineering background, they pointed out specific functions that could provide important support to this task :\\

\checkmark \textit{DS11: "I really miss writing unit testing and being able to lint code. The Jupyter notebook does not have linting, everyone writes the code they want, and today we have tools, such as Black, Isort, Pylint, Flake8, Bandit, and it is very difficult for you to use them in Jupyter notebooks. I think this lack of lint, this lack of testing is crucial for me."}\\

The difficulty of testing and debugging the code can influence the data scientist's ability to identify or even fix a bug, which may affect the number of accepted answers (the answer that solved the problem), the number of unanswered posts, or the acceptance time of an answer in StackOverflow.

\begin{figure}[h]
\includegraphics[width=10cm]{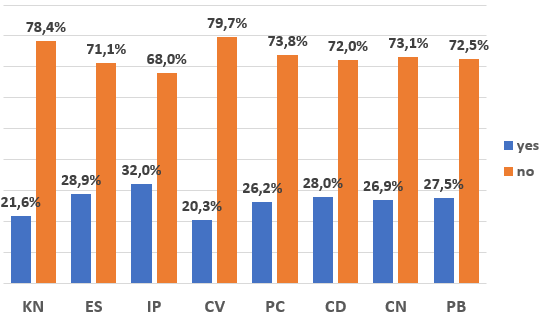}
\caption{Bug reports with accepted answers}
\label{accepted_answer}
\end{figure}

\begin{figure}[h]
\includegraphics[width=12cm]{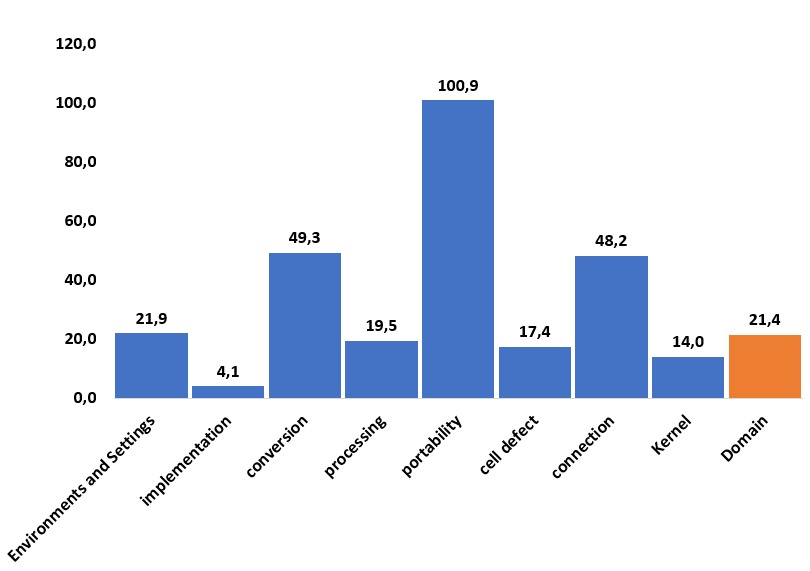}
\caption{Average accepted response time (in days) }
\label{accepted_time}
\end{figure}

Figure \ref{accepted_answer} shows the number of questions reported with the accepted answer, considering each bug type in our StackOverflow dataset. All the bug types have a similar average (27.9\%) of accepted answers. Figure \ref{accepted_time} shows the average time to get an acceptable answer. The average time to obtain an acceptable answer in the Jupyter notebooks domain is 21 days, at least 4 out of 8 bug types are above average.

\begin{mybox}{Finding 4}
Data scientists perceive bugs differently. Their hands-on experience with software engineering techniques can change how they identify bugs. In addition, the lack of basic features in Jupyter to promote code testing and debugging can generate difficulties in fixing bugs.

\end{mybox}

\textbf{Data analysis deployment. } Jupyter notebooks are used in two distinct scenarios: first, the notebook itself is a product and it is further used to replicate or perform new analyses; and next, it can be encapsulated and added to another system to use it \cite{DBLP:conf/chi/ChattopadhyayPH20}. Some interviewees (DS3, DS5, DS7, DS8, DS9, DS11, DS12, DS14, DS15, DS16, DS17, DS18) reported that Jupyter notebook is a good tool for exploratory analysis and prototyping, but it has some limitations, such as the lack of basic features that could help convert notebook code or facilitate this process when generating a final product to be deployed:\\

\checkmark \textit{DS11: "It has some issues, especially if you want to generate a deliverable of what you are doing inside a Jupyter notebook."}\\

\checkmark \textit{DS7: "You can opt for a workaround, but it's not trivial when you are dealing with libraries you build, functions you build, or classes. (...) How do you matter, how do you build this environment, where you have solutions that use a library you created, for example. Maybe if Jupiter itself helped the user to already build the entire class structure and the entire code structure, or even if it offers tools to facilitate things like encapsulating a library, it could be something interesting too."}\\

Many interviewees use Jupyter notebook in industrial and robust projects and deploy it inside company systems to perform the analysis:\\

\checkmark \textit{DS7: "When you intend to deploy the Jupyter notebook code inside the system code, it is not a trivial task since some changes need to be performed to be properly deployed in production environment. It would be interesting if the Jupyter Notebook provide tools to support this deployment."}\\

\textbf{Bad Programming Practices. } Except for DS1, all other interviewees whose academic background was not computer science started in the field of data science and/or programming in Python through Jupyter, which raises the concern about what culture of code quality in computational notebooks is being propagated..

\begin{table}[ht!]
\begin{tabular}{ p{0,5cm} p{0,5cm} p{0,5cm} p{0,5cm} p{0,5cm} p{0,5cm} p{0,5cm} p{0,5cm} p{0,5cm} p{0,5cm}}
\hline
 \textbf{} & \textbf{2014} & \textbf{2015} & \textbf{2016} & \textbf{2017} & \textbf{2018} & \textbf{2019} & \textbf{2020} & \textbf{2021} & \textbf{2022}\\
 \hline
Posts & 37 & 381 & 1688 & 2559 & 3976 & 5490 & 7107 & 8607 & 531\\
Bugs & 9 & 121 & 663 & 1054 & 1712 & 2482 & 3378 & 3879 & 251\\
 \hline 
\end{tabular}
\caption{\label{Jupyter_history} Jupyter History - Posts vs Bugs}
\end{table}

The Jupyter notebook appears in StackOverflow annual survey\footnote{https://insights.StackOverflow.com/survey} since 2017, and as shown in Table \ref{Jupyter_history}, the posts and bugs on StackOverflow only increased. It reinforces the importance of evolving the tool with features to mitigate bugs and help data scientists to do exploratory analysis, prototyping, computational narratives, and generate products without losing quality. These aspects were also highlighted by a professional during our interviews:\\

\checkmark \textit{ DS11: "Another mistake is also... Generally when we write a Jupyter notebook we do not care much about the quality of the code, we write code in almost any way, we do not care about a Lint, things like that, right. People do not bother to test too, so I think it is one of Jupyter biggest problems. We do not appreciate our code, we do not care much about code quality, we do not care much about unit tests."\\}

\begin{mybox}{Finding 5}
Transforming an analysis developed in Jupyter into a product can be one of the most important features for data scientists in the industry. However, there is still a lack of resources to improve the code quality and this transition process. Some users have been looking for alternative solutions that combine the benefits of a Jupyter notebook and an IDE. The lack of resources focused on code quality can also lead new data scientists to have bad programming habits.
\end{mybox}

\section{Discussion}
In this section, we discuss the implications of our study's results. In particular, the implication for tool builders, researchers, and data scientists.

As highlighted in Finding 1, the most frequent bugs in the Jupyter notebook are related to Environments and Settings, consisting of 43.2\% of analyzed posts from StackOverflow and 35.5\% of the issues analyzed in GitHub. Majority of the root causes of this bug category were configuration issues, version issues, and deprecation-related issues, suggesting that a significant amount of time and effort expended by data scientists are spent dealing with these issues. If software-engineering research can aid data scientists, this would potentially save a substantial amount of time.

Incorrect algorithm implementations cause many bugs (44.2\% of GitHub issues and 22\% of StackOverflow posts). Most of them are related to coding and logical error resulting in "Incorrect Functionality" (Table 4). We posit that this is happening due to data scientists not being familiar with the existing software quality assurance techniques such as unit testing,  bug localization, and repair. Our intuition is corroborated by the findings reported in Finding 4. This calls for action from the software engineering community researchers and practitioners alike to increase awareness about the existing techniques and make such tools available for data scientists. Also, researchers need to develop tools that can seamlessly integrate with the Jupyter notebook, making it easy for data scientists to adopt the techniques.

Our study highlights lack of functionalities that are standard practice in Software Engineering. For instance, Version control systems (i.e., Git) are standard tools used in software development. However, in Jupyter notebook development, it is not standard practice yet, as mentioned by interviewee DS7, DS11 previously. Since existing version control systems do not compare differences in the generated Graphical User Interface (GUI) components, it is difficult to identify the differences between GUI components across different versions of a given notebook. So a tool helping developers to compare  GUI changes instead of only textual changes can help Jupyter notebook developers significantly. Interviewees also highlighted the lack of functionality to preview, explore and interact with the raw dataset before starting analysis and modeling, which can be a better alternative than the notebook cell visualization. Another common feature requested by interviewees was advanced debugging capabilities such as a viewer of the variables defined in the notebook and the values assigned in each cell. We posit that such easy-to-use debugging capabilities will help reduce the significant number of implementation errors in Jupyter notebooks. 

In our study, we noticed that Jupyter Notebook is a very useful solution when it comes to analyzing, investigating and exploring data, 95\% of our respondents reported understanding that the main (or only) usefulness is in these steps, as in contrast to other traditional IDEs like R-Studio or VS-Code its simple layout facilitates and highlights the analysis performed. Although almost all respondents reported this tool's potential in data exploration, 79\% of them reported a difficulty in transforming the analysis done in the Jupyter Notebook into code to be put into production and the lack of features that facilitate cleaning and adaptation of the code for transposition.

Finally, with the analysis of bugs and interviews, we brought a non-exhaustive list (see Table \ref{features}) of features desired by users. Some features already have ready-made extensions, but in our analysis the use of some extensions is not trivial, in addition to generating compatibility, version and configuration errors. That's why it's important to have an extension with a unified package of solutions for Jupyter Notebook or that some of these solutions are in the standard version of the tool.

\begin{table}[H]

\begin{tabular}{ p{4,5cm} p{8,0cm}}
\hline
\textbf{Feature} & \textbf{Description}\\
\hline
\multirow{ 2}{*}{Indentation corrector} & For Python indentation is important. Jupyter allows indentations to be the developer's responsibility while writing. The indentation corrector identifies and corrects wrong indentations at development time.\\

\multirow{ 2}{*}{Syntax highlighting} & Function that inspects the code indicating syntax errors, structure errors, etc.\\

\multirow{ 2}{*}{Data Preview} & Functionality to preview and explore the raw dataset before starting analysis and modeling, a better alternative than the notebook cell visualization\\

\multirow{ 2}{*}{Graphic Interaction} & Functionality to manually interact with the graphs generated during data analysis\\

\multirow{ 2}{*}{Multi-Languages Per Cell} & Possibility to use other programming languages in the same notebook\\

Version control & Notebook change manager\\
Text Editor & More advanced code editing features\\
\multirow{ 2}{*}{Development Framework} & Framework that provides a base architecture adapted for the notebook\\

\multirow{ 2}{*}{Real Time Collaboration} & Functionality to support people working together at the same time, even if they are in different places.\\

\multirow{ 2}{*}{Variable Manager} & Viewer of the variables defined in the notebook and the values assigned in each cell\\

\multirow{ 2}{*}{Connection Between Notebooks} & Functionality for the user to visualize his set of notebooks and make calls to notebooks and cells external to the current notebook.\\
\hline 
\end{tabular}
\caption{\label{features} Features mentioned by respondents}
\end{table}

\section{Threats to Validity}

In this section, we discuss several threats to validity for our study.

\textbf{Projects Selection.} We have not analyzed proprietary repositories, and our findings are limited to open source projects which may not be representative and comprehensive. We mitigate this limitation by mining a large number of 105 open source projects from GitHub selected based on a well-defined set of criteria.

\textbf{Bug Selection.} {We only collected the issues with a set of keywords in the commit message (see Section 2.1.). Even with a predefined list used also in previous research \cite{DBLP:conf/icse/GarciaF0AXC20, DBLP:conf/icse/Makhshari021}, it is possible to miss some real bugs that do not have these keywords}.   

\textbf{Manual Analysis of Bugs.} Our study involved manual inspection of bugs which is a potential error-prone process. In order to mitigate this threat, three authors (2nd, 3rd, and 4th) analyzed the bugs separately. Next, all divergence in the process was discussed with the whole team until a consensus was reached. Our results are also online for public scrutiny. 

\textbf{Quality of posts.} The trustworthiness of the posts collected from Stack Overflow can be a threat to our study. To mitigate this threat, we used an approach similar to \cite{10.1145/3338906.3338955} which collected the posts based on a score at least 5 and reputation of users asking the questions. This score can be used as a good indicator to trust the post as a good discussion topic among the developers' community that cannot merely be solved using an internet search. In addition, the reputation of the users asking questions on Stack Overflow can be a threat to the quality of the posts. We only investigated top scored posts which are from users with different range of reputation ranging from novices to experts.

\textbf{Taxonomy.} The final taxonomy is not fully comprehensive since the derivation process is dependent on the collected commits, posts on Stack Overflow, and authors judgment. We mitigated this threat investigating 105 Jupyter notebook repositories, 14740 valid commits, and 30416 Stack Overflow posts. 

\textbf{Interviews.} The interviews were conducted with open-ended questions, where the participants were asked to express their perceptions and point-of-views. The interviews were conducted at 12 different companies and when these happened in the same company, the participants were warned not talk to each other about it to avoid bias. In addition, we did our best to select experienced professionals at each company to avoid our sample not being mature enough to have the expressive knowledge about our area of investigation. Another aspect that is critical for validity is the quality of the material used in the study. Thus, to ensure that the interview prompt had high quality, a pilot interview was conducted with a professional data scientist. Finally, to avoid the threat of concluding false conclusions based on the interview data, we carefully validated our interviews and findings with the participants as we performed analysis, sometimes asking for clarifications. 

\section{Related Work}

In this section, we discuss the main work related to our study.

\textbf{Jupyter Notebooks - Extensions. }
The Jupyter Notebook Project aims to provide to the data science community, a simple graphic interface to promote the computational narrative, based on usability, collaboration and portability \cite{ProjectJupyter}. Some studies have been proposing different ways to improve these aspects.

Rule et al. \cite{10.1145/3274419} investigated how cell folding can contribute to notebook navigation and reading. They developed an extension for it, but in some cases, folded sections were ignored or increased the time of notebook revisions. It shows how the analysis process in a notebook can be confusing and hard to understand, especially in large documents. Head et al. \cite{DBLP:conf/chi/HeadHBDD19} developed a solution to collect and organize code versions, helping the analyst to study, review and recover old codes and analysis. 

Computer notebooks unify text, code and visual outputs, being able to manually interact with the graphical outputs increases the data analysis power of scientists. Kery et al. \cite{DBLP:conf/uist/KeryRHMWP20} developed an API for this. 

Considering to support reproducibility, Wang et al. carried out two studies. The first one to recover the notebook's reproducibility with a tool that generates possible execution schemes \cite{10.1145/3324884.3416585}, and the second one to retrieve and install the notebook's experimental dependencies \cite{DBLP:conf/icse/WangLZ21}.

Our study is not focused on producing new features for Jupyter Notebooks. We analyze, identify and classify bugs in the Jupyter notebook to provide a systematic overview of bugs and developer challenges and provide an initial body of knowledge for future work on gaps and limitations in the daily use of the Jupyter notebook.

\textbf{Jupyter Notebooks - How Data Scientists Use. }
Some studies explore how the data scientists use the notebooks in their daily usage. Code duplication, for example is a common practice from data scientists. Koenzen et al. \cite{DBLP:conf/vl/KoenzenES20} studied how these duplication happen and identified that although there is an approximately 8\% rate of duplicate code in GitHub databases, users prefer not to duplicate their own code. 

Data analysis processes provide insights that need to be demonstrated, shared and disseminated. Wang et al. \cite{10.1145/3359141} studied the real-time collaboration and identified that working on synchronous notebooks encourages exploration and reduces communication costs, but the resources currently available for this imply the need for greater team coordination.

\textbf{Jupyter Notebook - Notebook Quality. }
Chattopadhyay et al. conducted a study \cite{DBLP:conf/chi/ChattopadhyayPH20} that involved observing five data scientists at their work with computational notebooks. They interviewed 15 data scientists and next surveyed 156 data scientists. They cataloged nine main problems and difficulties faced by data scientists using computer notebooks. Unlike this research, our study highlights the challenges faced by users from the perspective of real bugs that data scientists encounter in their daily work.

Rule et al. \cite{10.1145/3173574.3173606} analyzed the structure of 1 million notebooks to assess whether they were being built to support computational narratives. They identified that most of the notebooks are built without proper cleaning or documentation, making readability, replication, code reuse and consequently reproducibility a difficult task. Pimentel et al. \cite{DBLP:conf/msr/PimentelMBF19} conducted a large-scale study on notebook reproducibility problems. Their results show that only 24.11\% of notebooks run without errors, and out of that percentage, only 4.03\% are able to produce the original results. Later, they conducted another study that conducted a more detailed analysis \cite{PimentelMBF21}. While the authors are interested in analyzing notebooks regarding their structure, our study aims to understand the notebook code quality throughout the existing bugs.

Investigating the coding quality of Jupyter notebooks, Wang et al. \cite{10.1145/3377816.3381724} developed a preliminary study where the results revealed a high amount of bad coding practices in Jupyter notebooks. However, unlike the previous study, Patra et al. \cite{DBLP:journals/corr/abs-2112-06186} decided to focus on a single type of coding inconsistency that appears in Jupyter notebooks, Name-Value, and its implications for understanding and maintaining code. Unlike the previous studies that cite specific bugs, our work categorize and quantify types of bugs and root causes in the domain of Jupyter notebooks.

\textbf{Empirical Studies on Bugs.}
Some related work are not directly related to data science, such as: Zhang et al. \cite{10.1145/3213846.3213866} mined bugs in deep learning applications based on Tensorflow. They analyzed GitHub commits, pull requests and issues and StackOverflow questions. Using similar mining strategies and same data sources, Islam et al. \cite{10.1145/3338906.3338955} extended the search for other popular deep learning libraries, Caffe, Keras, Tensorflow, Theano, and Torch. In addition, Thung et al. \cite{DBLP:conf/issre/ThungWLJ12} analyzed bugs in machine learning systems, but their research used the issues reported on Jira database as a data source. However, to the best of our knowledge, this is the first empirical study of bugs in Jupyter Notebook projects.

Other studies focused on bugs by only analyzing the GitHub projects in different domains, such as bugs in autopilot software in unmanned aerial vehicles \cite{10.1145/3468264.3468559}, bugs in IoT systems \cite{DBLP:conf/icse/Makhshari021}, bugs in autonomous vehicles \cite{DBLP:conf/icse/GarciaF0AXC20} and bugs involving Infrastructure as Code Scripts \cite{10.1145/3377811.3380409}. 

All previous research focused on analyzing specific aspects of bugs, such as symptoms, commonality, bug evolution, bug prone stages, and bug detection. Our work is a preliminary study that focuses on providing the characterization of bugs in Jupyter Notebook projects, such as the types of bugs, the potential root causes, their frequency, and the impact and challenges for data scientists.

\section{Conclusion}
In this work, we conducted a large-scale empirical study to characterize bugs in Jupyter notebook projects. First, we analyzed 855 commits from 105 GitHub open-source repositories. Next, we analyzed 2585 Stack Overflow posts which gave us insights into bugs that data scientists face when developing Jupyter notebook projects. Finally, we conducted semi-structured interviews with 19 data scientists from 12 companies to validate the findings. We proposed a taxonomy of Jupyter notebook-specific bugs by analyzing these bugs. In particular, we identify eight classes of bugs, ten types of root causes, and the impact of bugs. 

The most frequent bugs in the Jupyter notebook are those related to Environments and Settings and Implementation. Regarding the root causes, the most frequent were: Configuration issues, Version issues, and Coding Errors. They are the cause of most Implementation and Environments and Settings bugs. The most frequent bug impact was Run Time Error, followed by Incorrect Functionality. In addition, we found that the data scientist's background determines how the bugs are identified, highlighting the importance of testing and debugging tools. Finally, we identified the Jupyter notebook deployment as a challenging and poorly supported task.

We believe this study can facilitate practitioners' understanding of the nature of bugs and define possible strategies to mitigate them. Our findings can guide future research in related areas, such as developing tools for detecting and recommending fixes for bugs in the Jupyter notebook.

\bibliographystyle{ACM-Reference-Format}

\end{document}